\documentclass[sigconf,nonacm]{acmart}

\AtBeginDocument{%
  \providecommand\BibTeX{{%
    \normalfont B\kern-0.5em{\scshape i\kern-0.25em b}\kern-0.8em\TeX}}}

\usepackage{graphicx}
\usepackage{balance}  

\usepackage{xcolor}
\usepackage{enumitem}
\usepackage{amssymb}

\usepackage{url}
\usepackage{array}

\newcommand{\nop}[1]{{}} 

\newcommand{\pl}{\emph{S2CE}}
\newcommand*\rot{\rotatebox{90}}

\setcopyright{rightsretained}
\copyrightyear{2020}
\acmYear{2020}
\acmDOI{}

\acmConference[]{}{}{}
\acmISBN{}

\begin{document}

\title{S2CE: A Hybrid Cloud and Edge Orchestrator for Mining Exascale Distributed Streams}


\author{Nicolas Kourtellis}
\email{nicolas.kourtellis@telefonica.com}
\affiliation{%
	\institution{Telefonica Research}
	\city{Barcelona}
	\country{Spain}
}

\author{Herodotos Herodotou}
\email{herodotos.herodotou@cut.ac.cy}
\affiliation{%
	\institution{Cyprus University of Technology}
	\city{Limassol}
	\country{Cyprus}
}

\author{Maciej Grzenda}
\email{m.grzenda@mini.pw.edu.pl}
\affiliation{%
	\institution{Warsaw University of Technology}
	\city{Warsaw}
	\country{Poland}
}

\author{Piotr Wawrzyniak}
\email{800383@edu.p.lodz.pl}
\affiliation{%
	\institution{Lodz University of Technology}
	\city{Lodz}
	\country{Poland}
}

\author{Albert Bifet}
\email{albert.bifet@telecom-paris.fr}
\affiliation{%
	\institution{LTCI, Telecom Paris, IP-Paris}
	\city{Paris}
	\country{France}
}

\renewcommand{\shortauthors}{N. Kourtellis, et al.}

\widowpenalty10000
\clubpenalty10000

\begin{abstract}
The explosive increase in volume, velocity, variety, and veracity of data generated by distributed and heterogeneous nodes such as IoT and other devices, continuously challenge the state of art in big data processing platforms and mining techniques.
Consequently, it reveals an urgent need to address the ever-growing gap between this expected exascale data generation and the extraction of insights from these data.
To address this need, this paper proposes \textit{Stream to Cloud \& Edge} (\pl), a first of its kind, optimized, multi-cloud and edge orchestrator, easily configurable, scalable, and extensible.
\pl\ will enable machine and deep learning over voluminous and heterogeneous data streams running on hybrid cloud and edge settings, while offering the necessary functionalities for practical and scalable processing: data fusion and preprocessing, sampling and synthetic stream generation, cloud and edge smart resource management, and distributed processing.
\end{abstract}



\maketitle

\newpage 

\section{Introduction}
\label{sec:introduction}

\begin{figure*}[t]
	\begin{center}
		\includegraphics[scale=0.54]{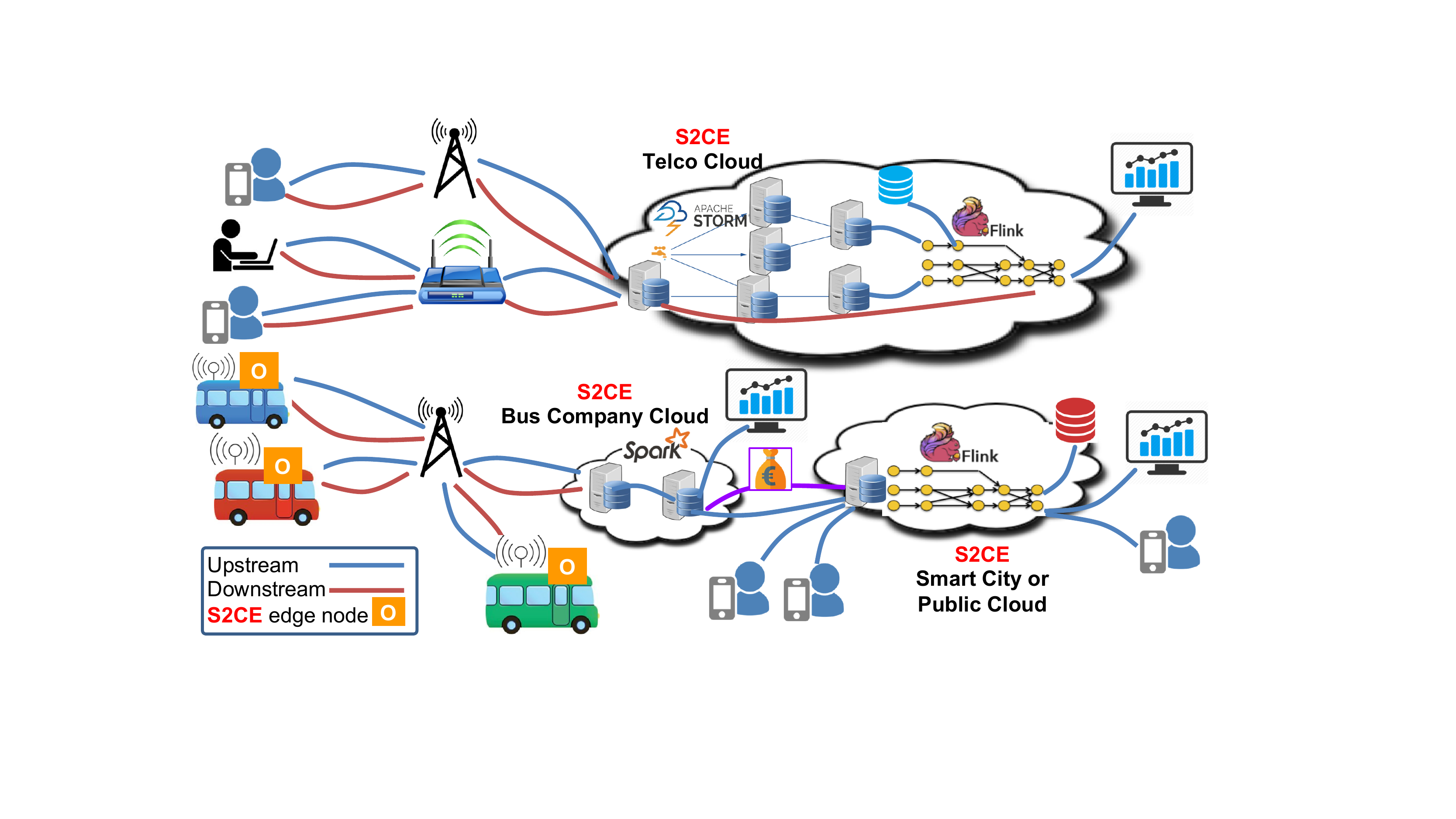}
		\caption{Examples of functionalities and usage of the \pl\ platform ecosystem.}
		\Description[]{Examples of functionalities and usage of the \pl\ platform ecosystem.}
		\label{fig:motivating-scenarios}
	\end{center}
\vspace{-4pt}
\end{figure*}

In the future Internet era, with hundreds of billions of devices, principle factors dominating the continuous utility of the Internet will be:
1) the massive population of devices and their intelligent agents,
2) the big, fast, and diverse data produced from them and their users, and
3) the need for large-scale, adaptive infrastructures to process and extract knowledge from these exascale data in order to help make critical, data-driven decisions.
Intelligent agents already exist in different forms, and are well embedded in various ways in our everyday lives, either as passive data collectors, or active producers.
They are instantiated in self-driving vehicles~\cite{self-driving-jat14}, phones~\cite{appdialogue-lrec16}, IoT devices~\cite{ai-iot-cs18}, personal artificial intelligence (AI) assistants~\cite{personal-assistant-jhs16}, factory or health sensors~\cite{ai-health-care-air16}, smart utility meters~\cite{smart-grid-survey16}, sensors in public transportation vehicles~\cite{ai-transport-jtl15}, chat bots~\cite{chatbots-asq17}, etc.
Such entities typically interface with centralized cloud processing systems, responsible for collecting, analyzing, and visualizing the data produced.
Finally, through machine learning (ML) and deep learning (DL), they can collect and learn new modalities of data, build new models and functionalities, and operate autonomously, making data-driven, critical decisions.

Human and smart agents' activities already produce big data workloads every day.
Billions of messages per day are processed by Facebook Messenger and WhatsApp~\cite{Hem16}, while the Internet of Things (IoT) ($\sim$75 billion objects by 2025~\cite{SIOT}) continuously produces data without human intervention, leading to a dramatic increase of data volume and velocity.
These numbers are only expected to exponentially grow through time: billions of devices (autonomous agents or user-devices) will generate big data continuously, as a stream, characterized by at least 4 important dimensions: Volume, Velocity, Variety and Veracity.

All these data are useful only when processed and modeled, and learnings are extracted in time to be used for appropriate decisions.
Thus, in the last decade there has been a high demand for data mining tools that allow data practitioners to compute complex ML models on big data streams produced by different sources and collected in centralized locations for processing.
To partially satisfy the need for computation power to process these data, we have witnessed an exponential growth of cloud computing, where the market itself is expected to reach 150 billion USD by the early 2020s~\cite{SHCC}.
Various vendors have introduced different types of clouds (private, public, and hybrid), with heterogeneous resources available in each, varying with respect to computation (CPU/GPU), memory, and network capacities.
However, all such clouds are typically not interoperable and do not facilitate data or computation portability between them.
Thus, customers are typically stuck in vendor lock-in, forced to rely on data analytic services of a single cloud provider, leading to lost opportunities in revenue and business from both sides (customers and providers)~\cite{martins16vendor-lockin}.
Finally, customers who choose private clouds are typically forced to deal with significant overhead in setting, managing, and manually tuning cloud resources.

Furthermore, the current model of data collection and processing is not sustainable.
Billions of IoT/edge devices generate tens of times more data than the 30+ million nodes across public and private cloud centers~\cite{Sha18}.
This influx of data cannot be processed in real time due to latency issues, lack of scalability in some cases, limited bandwidth in wireless connections (e.g., rural or congested areas), or compliance and privacy on data access, and does not allow customers to make real-time, data-driven decisions for their businesses.
The cloud industry has recently moved into a hybrid of cloud-edge computing, but this creates a whole set of new challenges regarding interoperability and APIs, managing heterogeneous capacities, workload offloading, data integrity and privacy, storage decentralization, application restructuring, etc.~\cite{Sco18}.

\vspace{2pt}
\noindent
\textbf{Proposal.}
To address the aforementioned challenges, this paper proposes \textit{Stream to Cloud \& Edge} (\pl), a first of its kind, optimized, multi-cloud and edge orchestrator.
In fact, \pl\ enables machine and deep learning on big data streams using hybrid cloud and edge resources, while offering the necessary functionalities for practical and scalable event-based processing: data fusion and preprocessing, sampling and synthetic data stream generation, cloud and edge smart resource management, and distributed event processing. %

\vspace{2pt}
\noindent
\textbf{Motivating scenarios.}
\label{sec:scenarios}
Future industrial settings can impose potentially diverse and interdisciplinary constraints and requirements to \pl\ and its tools.
However, its unique architectural design and functionalities enable it to flexibly accommodate different future use-case scenarios.
Figure~\ref{fig:motivating-scenarios} illustrates some of these functionalities and industrial setups.
\pl\ can consume big data from different sources, destined for different types of analysis.
These input data can range from data-in-motion, such as events produced from sensor and other readings from mobile or IoT edge devices, up to fully processed data-at-rest or in-motion from other complete cloud platforms.
The distributed and parallel nature of \pl's components allow the platform to scale gracefully to accommodate future exascale volumes and velocities.

Depending on the capabilities of the source nodes, pre-models could be executed at the edge to alleviate computing pressure from the main cloud platform.
Results can be consumed by end-users, the company managing the \pl\ platform, or shared and even sold to other downstream companies.
In fact, interconnection APIs made available by the platform can enable such data to be shared across companies, with appropriate payment schemes in place, while respecting users' privacy.
For example, a bus company may be collecting data to optimize bus routes and perform data-driven business decisions, but can also sell such data to a City Council to bootstrap their effort to optimize traffic, and reduce congestion and pollution in the city.
This crucial property will drive high business innovation in the data sharing markets, as required and expected by the emerging world Data Economy.

\vspace{2pt}
\noindent
\textbf{Contributions.}
With this paper, we make the following contributions in the domain of distributed, event-based processing systems:
\begin{itemize}[leftmargin=*]
\item Analyze state-of-the-art stream data processing methods, libraries, and systems that are widely used in academia and industry.
\item Identify needs and challenges faced by -- as well as success criteria expected by -- the industrial and R\&D sectors.
\item Propose a novel, hybrid, cloud-edge architecture that can address these challenges, with several key design objectives in mind.
\item Discuss the innovation potential of the proposed architecture across different dimensions.
\end{itemize}

\section{State-of-the-Art Analysis}
\label{sec:related}

The current state-of-the-art in the space of distributed data stream processing systems is populous and covers various aspects of the needs that the industry has from such systems.
In the next paragraphs, we briefly cover efforts from academia and industry to address such needs, and identify pending issues and gaps that a new platform, such as our proposed \pl, should address:
\begin{itemize}
\item Big data stream processing systems (Section~\ref{sec:related:systems})
\item Cloud resource management and tuning (Section~\ref{sec:related:cloud})
\item Distributed stream processing at the edge (Section~\ref{sec:related:edge})
\item Machine and deep learning over data streams (Section~\ref{sec:related:ml})
\item Data transformation techniques (Section~\ref{sec:related:transform})
\end{itemize}

\subsection{Big Data Stream Processing Systems}
\label{sec:related:systems}

There are currently several open source Distributed Stream Processing Engines (DSPE), such as Apache Storm~\cite{storm-website}, Apache Samza~\cite{samza-website}, and Heron~\cite{heron-sigmod15}.
While they support developing ML applications, they do not have dedicated ML libraries.
Apache Spark~\cite{spark-website}, originally a batch processing framework, now offers streaming support for micro-batches or continuous streams.
Spark ML and MLLib are Spark's ML libraries with only simple linear classifiers and clustering algorithms for streaming, none of which are state of the art.
Also, StreamDM~\cite{streamdm} is an Apache-licensed, open source software for mining big data streams using Spark Streaming, and developed by Huawei.
Apache Flink~\cite{flink-website} is a processing framework that focuses on streaming tasks and arranges them in directed acyclic graphs, similar to Spark's topologies.
Its ML library, FlinkML, supports batch-based ML.
Apache Apex~\cite{apex-website} and Apache Beam~\cite{beam-website} are unified stream and batch processing engines containing basic ML libraries.

In the Cloud front, Google Cloud Dataflow~\cite{gcd-website} is a streaming data processing system that can emulate batch processing and has some support for ML.
MS Azure~\cite{ms-azure-website} is Microsoft's cloud system performing both batch and stream processing. The batch processing engine has ML algorithms, but the Streaming Analytics tool can only compute aggregation and statistics, without any other streaming ML.
AWS Kinesis~\cite{aws-kinesis-website} is Amazon's offering for processing data streams in real-time but, similar to MS Azure, it does not provide native support for streaming ML.

Massive Online Analysis (MOA)~\cite{moa-jmlr10} consists of well-known online algorithms for streaming classification, clustering, and change detection mechanisms.
However, MOA only runs in a single machine and lacks high-performance integration interfaces (e.g., with the widely-used Kafka), making it non-usable in industrial big data deployments.
Vowpal Wabbit~\cite{vowpal-wabbit-website} is a streaming ML framework based on the perceptron algorithm with a specific focus on reinforcement learning and optimized for a multi-core single-node setting.
Jubatus~\cite{jubatus-website} is a ML framework for stream mining that establishes tight coupling between the ML library and the underlying custom-built DSPE, which limits the framework's applicability and extensibility.
Finally, Apache SAMOA~\cite{samoa-website} allows for distributed computation of several ML algorithms over four DSPEs, namely Storm, Flink, Samza, and Apex.

A future processing platform should take advantage of existing engines (be it pure streaming or hybrid) for executing a given streaming ML task, while offering an API for extending ML algorithms available and runnable on the platform.
Table~\ref{table:feature_comparison} summarizes how the desired platform should differ from, and advance the features and functionalities provided by the most relevant prior research projects and large-scale data processing systems used by the industry today.

\begin{table}[t]
	\centering
	\caption{Feature comparison between the desired platform and related large-scale data processing tools and systems.}
	\label{table:feature_comparison}
	\vspace{-2pt}
\small
\setlength\tabcolsep{1.2pt}
\begin{tabular}{l*{13}c}
	\multicolumn{1}{b{4cm}}{
		{\scriptsize
		$\times$ \textit{means no support}\newline
		$~~!$ \textit{means partial support}\newline 
		$\checkmark$ \textit{means good/full support}\newline}
		\newline \vfill 
		Features/Capabilities}
	& \rot{Apache Storm} 
	& \rot{Apache Samza} 
	& \rot{Apache Spark} 
	& \rot{Apache Flink} 
	& \rot{Apache Apex} 
	& \rot{Apache Beam} 
	& \rot{Google CD} 
	& \rot{MS Azure ML} 
	& \rot{MOA} 
	& \rot{Vowpal Wabbit} 
	& \rot{Jubatus} 
	& \rot{Apache SAMOA} 
	& \rot{\textbf{Desired Platform}} \\
	\midrule
	Stream integration components  & $\checkmark$ & $\checkmark$ & $\checkmark$ & $\checkmark$ & $\checkmark$ & $\checkmark$ & $\times$  & $\times$  & $\times$ & $\times$ & $!$ & $\checkmark$ & $\checkmark$ \\
	Data preprocessing and fusion  & $!$ & $\times$ & $!$ & $!$ & $\times$ & $\times$ & $\times$ & $\times$ & $!$ & $!$ & $\checkmark$ & $!$ & $\checkmark$ \\
	Built-in synthetic data generator  & $\times$ & $\times$ & $\times$ & $\times$ & $!$ & $!$ & $\times$ & $\times$ & $!$ & $\times$ & $\times$ & $!$ & $\checkmark$ \\
	Stream-based machine learning & $!$ & $!$ & $!$ & $!$ & $!$ & $!$ & $!$ & $!$ & $\checkmark$ & $\checkmark$ & $\checkmark$ & $!$ & $\checkmark$ \\
	Stream-based deep learning    & $\times$ & $\times$ & $\times$ & $\times$ & $\times$ & $\times$ & $\times$ & $\times$ & $\times$ & $\times$ & $\times$ & $\times$ & $\checkmark$ \\
	Resource management    & $\checkmark$ & $!$ & $\checkmark$ & $\checkmark$ & $\checkmark$ & $\checkmark$ & $\times$ & $\times$ & $\times$ & $\times$ & $!$ & $!$ & $\checkmark$ \\
	Distributed platform    & $\checkmark$ & $\checkmark$ & $\checkmark$ & $\checkmark$ & $\checkmark$ & $\checkmark$ & $\checkmark$ & $\checkmark$ & $\times$ & $\checkmark$ & $\checkmark$ & $\checkmark$ & $\checkmark$ \\
	Open license (Apache preferred)    & $\checkmark$ & $\checkmark$ & $\checkmark$ & $\checkmark$ & $\checkmark$ & $\checkmark$ & $\times$ & $\times$ & $!$ & $!$ & $!$ & $\checkmark$ & $\checkmark$ \\
	\bottomrule
\end{tabular}

\end{table}

\subsection{Cloud Resource Management and Tuning}
\label{sec:related:cloud}

Cloud Compute and Storage services are traditionally utilized for creating and provisioning Virtual Machines (VMs) to host batch and stream processing applications~\cite{cloud-provision-survey-kis16}. 
This method offers a tight control of the infrastructure where the software is running, but comes at high cost of maintenance, as each machine has to be provisioned individually. 
When it comes to scalability and optimization, there is a strong need for careful capacity planning and manual intervention even with automated tools~\cite{docker-survey-jgc16}.
To alleviate these issues, the recent trend is to use containers instead of VMs as the minimal computation unit in the cloud. 
Docker makes it easy to generate and run such containers, while Docker Compose is a lightweight solution for orchestrating them~\cite{docker-website}.
For big workloads, Kubernetes~\cite{kubernetes-website} is a better fit for automating application deployment, scaling, and management.
Kubernetes supports scaline up or down a set of containers, but the decision has to be manually programmed. Also, when nodes fail or in overload cases, there is lack of automated tools for infrastructure management in a Kubernetes cluster. 

Data management systems have grown in scale, complexity, and number of installations~\cite{babu-db-mr-fntdb13}. Such systems contain 100s of configuration parameters and execute on 1000s of nodes that must be properly configured and managed~\cite{bi-big-data-17}. 
Hence, it is crucial to automate the process of resource management, as well as optimization and tuning of application performance.
The problem of resource allocation deals mainly with gathering and assigning resources to applications, while scheduling deals with allocating tasks to resources~\cite{survey-resource-alloc-icices14}. 
Past works focused on different aspects of the problem: 
\cite{stream-scheduling-sigmetrics15} investigates the scheduling problem satisfying load balancing and cost considerations;
\cite{stream-scheduling-cicis14} employs design-time knowledge and benchmarking method to deal with scheduling on heterogeneous clusters; 
Re-Stream~\cite{restream-is15} focuses on energy-efficient resource scheduling; 
\cite{tstorm-icdcs14} addresses adaptive scheduling. 
Others focus on automatically optimizing and tuning workloads using various techniques such as cost-based (e.g.,~\cite{starfish-cidr11, ernest-nsdi16, caladrius-heron-icde19}) or ML-based (e.g.,~\cite{rfhoc-tpds16, spark-ml-tuning-hpcc16, incremental-ml-tkde17}).

Overall, a future platform should employ new and advanced statistical and machine learning techniques for (i) understanding application behavior and cloud resource usage, (ii) optimizing the provisioning of cloud resources for applications across different providers in an automated and vendor-neutral manner, and (iii) automatically tuning applications to increase performance and meet SLAs.

\subsection{Distributed Stream Processing at the Edge}
\label{sec:related:edge}

Systems for distributed stream processing have traditionally been designed to run on clusters or in the cloud.
However, processing all the data there can introduce latency delays due to data transfer, which makes near real-time processing difficult to achieve. 
In contrast, edge computing has become an attractive solution for performing certain stream processing operations, and hence (i) reduce end-to-end latency and communication costs, (ii) enable services to react to events locally, or (iii) offload processing from the cloud~\cite{edge-5g-etsi15, spanedge-sec16}. 
Computing, storage, and network resources located at the network edge, however, are more constrained than those deployed in the cloud. 
While edge computing is often used to reduce latency of delivering content to mobile end-users, the emergence of application domains such as IoT require data events to be treated locally, under short time delays.

The deployment of data stream processing applications onto heterogeneous infrastructure has been proven to be NP-hard~\cite{BDN13}. 
Moreover, moving operators from cloud to edge devices is challenging due to limitations of edge devices~\cite{stream-edge-survey18}. 
Existing work often proposes placements strategies considering user intervention~\cite{spanedge-sec16}, whereas many models do not support memory and communication constraints~\cite{geelytics-bigdata16}. 
Studies also consider all data sinks to be in the cloud, with no feedback loop to actuators located at the edge~\cite{fog-allocation-iotj17}. 
Thus, no current solution covers scenarios involving smart cities, precision agriculture, and smart homes comprising heterogeneous sensors and actuators and time-constraint applications.

A future data stream processing service should be able to orchestrate the deployment of processing tasks and achieve resource elasticity under highly distributed environments comprising edge computing and clouds.
A lightweight version of the platform should be able to take advantage of existing solutions such as Apache Edgent~\cite{edgent-website}
in order to decentralize the online ML and mining towards the edge of the ecosystem.

\subsection{Machine/Deep Learning over Data Streams}
\label{sec:related:ml}

To effectively deal with streaming data, models must be able to adapt to patterns evolving over time by detecting changes in a fast and accurate way~\cite{concept-drift-survey14}. 
Thus, shift detection mechanisms are necessary in the context of devised ML methods, to render them self-adaptive, similar to DDM~\cite{ddm-bsai04}, EDDM~\cite{eddm-kdds06}, and ADWIN~\cite{adwin-acml10}.
Past work~\cite{visualization-is13, visualizing-asa15} also demonstrates the need to better understand the ML model structure and claim that such understanding is possible through informative visualization of the model structure. 
New sophisticated visualization approaches are the Forest Floor~\cite{forest-floor-arxiv16}, interaction importance extraction~\cite{ggrandomforests-arxiv15}, and the factorMerger~\cite{factorMerger-arxiv17}. 
Still the visualization of model structures is less mature than that of raw data provided by popular BI tools such as Tableau~\cite{tableau-website} and Power BI~\cite{powerbi-website}.
Hence, there is a strong need for designing and implementing adaptive distributed algorithms for learning from streaming data, novel tools for visualization of ML models, and tools for tracing and monitoring the changes in the evolution of streams and of the ML models.

Conventional deep networks do not allow for uncertainty representation, which is key for addressing learning from streaming data. 
Bayesian inference-based variants~\cite{baysian-ss09} offer a solution that can account for two types of uncertainty: 
(i) heteroscedastic uncertainty, which captures noise inherent in the observations due to temporal irregularities in data sampling; and 
(ii) model uncertainty, which accounts for uncertainty in the parameters. 
However, there are two main issues with using Bayesian updating on data streams.
First, Bayesian inference computes posterior uncertainty under the assumption that the model is correct, rather than being an approximation.
Second, existing approaches either explicitly model the time series, at the cost of low inferential performance, or assume that the data are exchangeable, i.e., that the underlying distribution does not change over time.

To address these issues, a future machine learning stream processing platform should adopt ideas from Bayesian non-parametrics, such as population-driven posteriors~\cite{population-posterior-nips15}, and introduce them into the configuration of postulated Bayesian deep networks, and adopt self-adaptive prior assumption mechanisms in the context of approximate variational inference.
These adaptations should enable the platform to build more complex, accurate ML and DL models.

\begin{table*}[t]
	\centering
	\caption{Expected industrial challenges and how the envisioned platform's design should address each challenge.}
	\label{tab:industrial-challenges}
\small
\setlength\tabcolsep{6pt}
\begin{tabular}{l l}
\toprule
Expected Industrial Challenge	&	(Objective) How does \pl\ address the challenge?			\\
\midrule
Heterogeneity	&	(O1) Handling diverse types of cloud computing resources				 \\
Scalability		&	(O1) Distributed and parallelized dynamic analytics for real-time learning		\\
Data-in-motion and data-at-rest		&	(O1) Processing data seamlessly at the same time without extra system overhead			\\
\midrule
Hybrid (central$+$edge) big data architectures	&	(O2) Optimizing an efficient mixture of central and edge resources	\\
Decentralization \& edge	&	(O2) Computing at edge for faster, more scalable, energy efficient processing	\\ 
\midrule
Data/AI/predictive/prescriptive analytics		&	(O3) Using distributed deep and machine learning		\\
Stream analytics frameworks \& processing	&	(O3) Minimal development effort, scalability, processing speed	\\
Advanced business analytics	&	(O3) Intelligence to empower companies for accurate, instant, data-driven decisions	\\
\midrule
Heterogeneity	&	(O4) Handling diverse data, modeling, and input/output interfaces	\\
Semantic interoperability	&	(O4) Facilitating data and model exchange between vertical data silos	\\
Data quality	&	(O4) Providing curation methods for data filtering, quality assessment, improvement	\\
Distributed trust infrastructures	&	(O4) Managing data in anonymized and decentralized fashion	\\ 
\bottomrule
\end{tabular}
\end{table*}

\subsection{Data Transformation Techniques}
\label{sec:related:transform}

Raw data often need to be transformed before processing. 
Typical architectures use Apache Flume~\cite{flume-website} or Kafka~\cite{kafka-website} to first capture data of interest from distributed sources, and then apply preprocessing such as filtering and format conversion.
Given the popularity of these systems, 
interfaces and methods of Kafka and Flume should be utilized for receiving input from multiple upstream sources and producing output consumable by downstream units. 
For streams with delayed labels, methods have been proposed such as
(i) drift detection applicable in stream classification~\cite{drift-detection-eais15}, 
(ii) classification method inspired by micro-clusters~\cite{delayed-labels-icmla15}, and 
(iii) analysis of upper loss bound of multiple expert system trained in nonstationary environment with verification latency~\cite{concept-drift-ijcnn14}.
Such methods are not available in DSPEs libraries such as Spark ML or FlinkML. 

Traditional dimensionality reduction techniques do not apply for stream data as the processed data arrive at real-time and the reduction must happen online, with no-multiple loop, batch-based algorithms. 
\cite{dim-reduction-stream-phd09} explored statistical inference methods for reducing dimensions of streams using hashing projections to derive efficient estimators of cardinality. 
\cite{dim-reduction-book13} discussed different subspace tracking methods for reducing dimension space in streams.

Studies like~\cite{synthetic-gen-kddm09, synthetic-gen-bigdata14} offered methods on how to produce synthetic streams from real data by inferring underlying statistical distributions. 
Such approaches do not work for streams with concept drifts, and protecting privacy and confidentiality cannot be done with fixed privacy preserving rules. 
Data generation allows altering volume, velocity, variety, and proportion of records linked to individual features, while maintaining inter-feature dependencies. 
Data generators in~\cite{data-gen-tcpeb10, datagenerator-website} allow for systematically producing large data volumes.
\cite{moa-jmlr10} offers several synthetic stream generators for benchmarking ML methods, but without capabilities to scale properly to produce large volume/velocity streams.

A future platform should provide instance data preparation for ML, data fusion, methods to deal with evolving, incomplete, or delayed data records and events, delayed labels, and time-spanned joins of streams.
Methods should also be explored based on newfound advancements in DL, capturing hidden similarities within streams, compressing more effectively data for efficient processing downstream.
Finally, such a platform should provide data generation process to 
(i) handle non-stationarity of data distributions due to changes in the environment, 
(ii) scale appropriately to produce required volume and velocity of data, 
(iii) handle concept drift and skewness observed in real data.

\section{Industrial Challenges \& Objectives}
\label{sec:needs}

In the next decade, the big data stream mining community will face several industrial challenges, as summarized in Table~\ref{tab:industrial-challenges}.
In fact, these challenges drive particular industrial needs, for which a future processing platform should adequately address.
In the next paragraphs, we first summarize four fundamental industrial needs stemming from our analysis of the current state-of-the-art, as well as success criteria that must be satisfied to address these needs.
Then, we outline the desired objectives that the future platform must have by design, to fulfill these needs and success criteria.

\subsection{Industrial Needs \& Success Criteria}

\vspace{2pt}
\noindent
\textbf{Industrial Needs.}
Based on the current challenges identified in Section~\ref{sec:related}, and the expected landscape on data generation and processing systems, we anticipate that the industry in data analytics on cloud infrastructures will have the following four future needs:
\begin{description}[leftmargin=*, itemsep=1pt]
\item[N1:] Computing platforms that can decentralize processing at the source of data during generation (i.e., edge), to alleviate pressure of computation and storage at cloud/centralized infrastructure.
\item[N2:] Computing platforms that use resources on heterogeneous multi-clouds (public/private) for data mining.
\item[N3:] Computing platforms that can automatically self-tune and orchestrate their resources for optimal resource allocation and use between cloud and edge.
\item[N4:] Advanced machine and deep learning tools, capable of preprocessing and analyzing exascale streams at real time, both at the edge where data are produced and at the cloud where more complex modeling can be done.
\end{description}

\vspace{2pt}
\noindent
\textbf{Success Criteria.}
A proposed platform can effectively address the above needs if it satisfies at least the following criteria:
\begin{description}[leftmargin=*, itemsep=1pt]
\item[S1:] The platform should scale appropriately and in a distributed fashion to sustain throughput while processing exascale data streams expected in the next 5-10 years.
\item[S2:] The platform should produce real-time data insights with microsecond updates, based on advanced machine and deep learning models computed on incoming streams.
\item[S3:] The platform should shift workload between cloud and edge resources seamlessly; increased latency and reduced model performance should not violate agreed SLAs.
\item[S4:] The platform should integrate fully with current and future big data processing systems, and facilitate easy adoption and usability by big data practitioners and engineers.
\end{description}

\begin{figure*}[t]
	\begin{center}
		\includegraphics[width=0.9\textwidth]{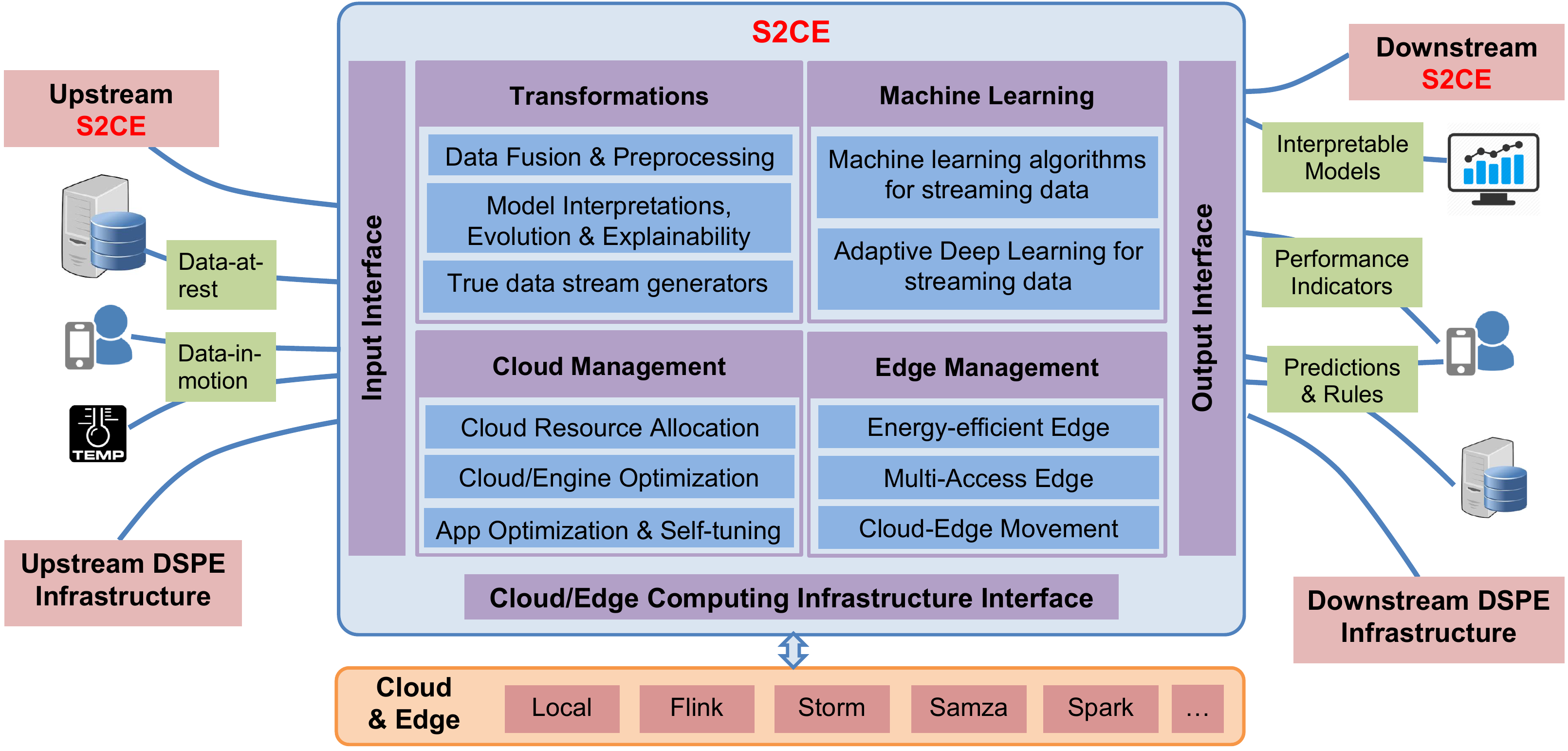}
		\caption{Global architecture of the proposed \pl\ platform.}
		\Description[]{Global architecture of the proposed \pl\ platform.}
		\label{fig:architecture}
	\end{center}
	\vspace{-8pt}
\end{figure*}

\subsection{Desired Platform Design Objectives}\label{sec:objectives}

The envisioned platform must have key properties embedded in its design to fulfill the following four main objectives, thus satisfying the future industrial needs of big data stream processing.
Table~\ref{tab:industrial-challenges} summarizes how each objective will address each industrial challenge outlined earlier.
In effect, the envisioned platform should:
\begin{description}[leftmargin=*, itemsep=1pt]
\item[Objective O1:] Be a data processing platform capable of scaling in a distributed fashion on cloud resources to ingest exascale streams, and utilize smart cloud resource management and self-tuning for reduced energy consumption, increased efficiency, easy configuration and maintenance.
\item[Objective O2:] Be a hybrid cloud-edge architecture capable of scaling in a decentralized fashion to preprocess big data streams on edge nodes, by utilizing smart edge resource management for workload migration from cloud to edge nodes, for energy efficiency and reduced latency.
\item[Objective O3:] Provide the next generation of advanced stream analytics, with distributed machine and deep learning (ML/DL), to support predictive and prescriptive analytics of both data-at-rest and data-in-motion in a unified manner, that empower business intelligence in the new paradigm of Data Economy.
\item[Objective O4:] Support tools for input/output data transformation and synthetic data stream generation based on real-world statistic distributions, for ensuring data quality, interoperability, and replicability, while enabling the privacy-preserving and confidential sharing of data.
\end{description}

\section{Platform Design}
\label{sec:overview}

Our proposed platform, \pl, is a complete, cloud-edge orchestration platform, based on four key components for input/output transformations, machine and deep learning processing, as well as cloud and edge resource management and tuning.
\pl\ aims to accelerate the building of tools that facilitate and enhance real-time artificial intelligence over voluminous and heterogeneous streams of data from IoT and other sources, in order to extract meaningful knowledge and perform extreme-scale predictive analytics over cloud and edge computing infrastructures.
A key objective of this platform is to provide the big data industry with standardized interconnection methods and a stream mining framework within the Apache big data ecosystem, to enhance and automate decision making processes.
In the next paragraphs, we provide an overview of the architectural components needed to instantiate the envisioned platform and its desired properties. 

\subsection{Architectural Components}
\label{sec:arch-components}

The high-level vision is to build an integrated platform that includes:
(i) input of data from different upstream infrastructures (these can be other \pl\ pipelines, edge components deployed on top of IoT or mobile nodes to perform data preprocessing, etc.); 
(ii) cloud- and edge-based modules combining ML algorithms and stream computing infrastructure; and 
(iii) output modules for data visualization or input to downstream pipelines.
Figure \ref{fig:architecture} shows the architecture of this conceptually novel platform with its various key components described in detail below.

\vspace{2pt}\noindent
\textbf{Input Interface:}
APIs providing standardized, secured interconnections will allow the mixing of multi-input data streams. 
Apart from the edge, such streams can come either directly from IoT-related sensors, or data producers including stream platforms positioned upstream in the pipeline. 
These data can be of different types (data-in-motion or data-at-rest), formats, and arriving at different velocities and volumes.
Due to its adaptive features described later, the platform will be capable of consuming them without penalties in performance (throughput, prediction accuracy, etc.), or security. 
It will also be capable of fusing, preprocessing, aggregating, or sampling diverse data.

\vspace{2pt}\noindent
\textbf{Transformations:}
The data streams will feed the Transformations component, with the following functionalities:
\begin{itemize}[leftmargin=*]
	\item \textit{Data preprocessing and fusion:} Data can be processed and transformed to improve the quality of data and learning.
	The transformations will be either instance- or attribute-based and will allow the imputation of missing data, fusing, and normalizing when multiple data types provided, and dealing with delayed data.
	This module will allow handling of complex, multi-featured data, with dimensionality reduction techniques either for machine learning or modeling within synthetic stream generators.
	\item \textit{Visual exploration and model explanation:} ML models built on top of streams are difficult to monitor their structure and performance.
	This module will contain methods for interpretable ML that efficiently summarizes and visualizes drift, model structure, evolution and performance.
	\item \textit{Changes in Online Models:} Changes in characteristics of data or models require human intervention and, if not addressed in time, could lead to model performance degradation.
	This module will also identify and visualize significant changes and trends in data and model performance.
	\item \textit{Privacy-preserving stream generators:} To test end-to-end platform performance and usability, novel synthetic data generators will be developed based on game neural networks.
	These generators can be used for sharing of synthetic data reflecting closed business data while preserving data owners' privacy, for benchmarking of both ML methods and end-to-end applications under varied load.
\end{itemize}

\vspace{2pt}\noindent
\textbf{Machine Learning:}
This component will take inputted data and perform various algorithms for fast learning. The main challenge will be that such algorithms need to be incremental, use a small amount of time and memory for processing, and adapt to the changes on the streams:
\begin{itemize}[leftmargin=*]
	\item \textit{ML streaming algorithms:} The platform will contain the necessary abstractions (and a library) so that complex ML algorithms can be implemented for classification, clustering, anomaly detection, frequent pattern mining, and reinforcement learning, designed to scale in a distributed fashion, using cloud deployments to consume very large data streams in real time.
	\item \textit{Self-adaptive DL algorithms:} DL (Deep Learning) algorithms that evolve and adapt on the streamed data, in a self-enforced fashion, will also be supported.
\end{itemize}

\vspace{2pt}\noindent
\textbf{Cloud Resource Management:}
This component will manage cloud resources efficiently, starting from the basic mechanics needed for cloud resource provisioning, to algorithmic monitoring and distributed task management, to self-tuning streaming applications automatically:
\begin{itemize}[leftmargin=*]
	\item \textit{Resource Allocation, Deployment \& Monitoring:} Fundamental methods will be deployed to allow \pl\ to monitor and control available resources at different cloud providers and allocate resources in them as needed, deploying computation tasks and monitoring execution.
	\item \textit{Cloud/Engine Algorithm Management:} Different cloud and computing engine parameters impact streaming applications in different ways.
	This module will be responsible for making initial and recurring provisioning and configuration decisions to meet performance, monetary budget, and/or energy efficiency objectives, while employing the most appropriate cloud and computing engine and, if needed, applying data transformation and reduction strategy satisfying these objectives.
	\item \textit{Optimization \& Self-Tuning of Cloud Applications:} Given a ML task to be performed on an input data stream, the platform will be able to self-tune using ML algorithms to pick the best streaming engine and appropriate parameter settings for the execution of the task.
\end{itemize}

\begin{figure*}[t]
	\begin{center}
		\includegraphics[scale=0.5]{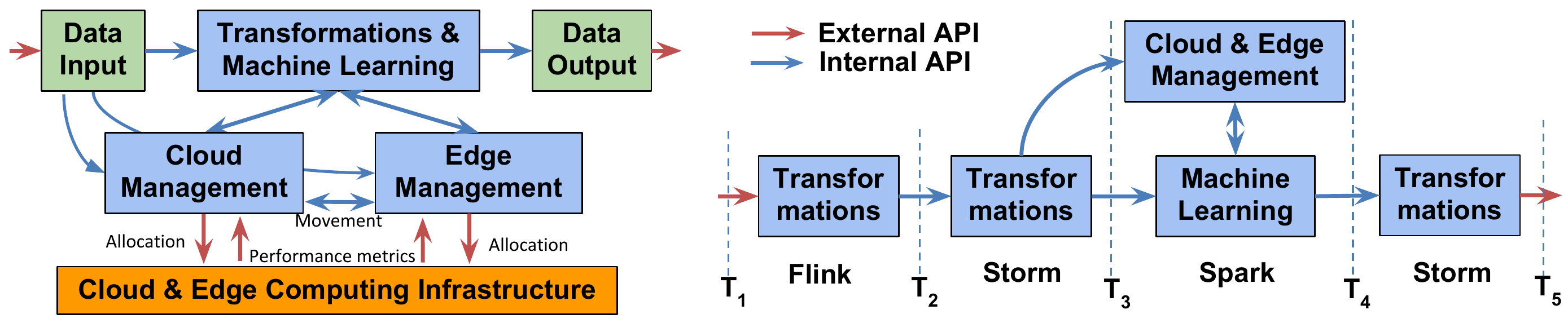}
		\caption{APIs and component interconnection. The components can run in one \pl\ instance (left), or multiple \pl\ instances and on different DSPE systems (right).}
		\Description[]{APIs and component interconnection. The components can run in one \pl\ instance (left), or multiple \pl\ instances and on different DSPE systems (right).}
		\label{fig:apis}
	\end{center}
	\vspace{-6pt}
\end{figure*}

\vspace{2pt}\noindent
\textbf{Edge Resource Management:}
This component will manage edge resources efficiently, by understanding what streaming computation can be offloaded from cloud to edge resources, where (i.e., to which edge nodes), when this offloading should be done (or reversed), as well as how the offloading will be done: 
\begin{itemize}[leftmargin=*]
	\item \textit{Energy-Efficient Edge Placement:} Many streams are simple enough to be preprocessed at their origin (edge) to reduce communication and processing costs at the main cloud platform. Sampling and summarization algorithms will be applied at the edge (e.g., IoT nodes, mobile devices), while guaranteeing property preservation of streams (e.g., via unbiased sampling), and dynamic reconfiguration of processing services according to computing and network availability of edge nodes.
	\item \textit{Multi-access Edge Computing (MEC):} The platform will contain a MEC module to enable collaborative deployment of applications on the basis of telecommunication and computing resources located closer to user equipment.
	\item \textit{Computation Movement between Cloud and Edge:} Fundamental blocks will be deployed to monitor utilization of the edge computing infrastructure and, in coordination with the Resource Allocation, Deployment \& Monitoring module, enable ML streaming algorithms to offload computation from cloud to edge resources and vice versa.
\end{itemize}

\vspace{2pt}\noindent
\textbf{Output Interface:}
This component will implement standardized, secured interconnections to allow the algorithmic results from the ML module to be outputted for downstream streaming engines, \pl\ instantiations, and even end-users to consume on their devices in a seamless, secure way through a pipeline.
\pl\ will offer new connectors and interfaces to create stream processing pipelines and output its resulting predictions and models for other engines to use downstream.
The output data streams will be splittable to different formats and substreams, depending on the utility of the overall pipeline.

\vspace{2pt}\noindent
\textbf{Computing Infrastructure Interface:}
This component will allow \pl\ to operate on top of several well-established distributed stream processing engines (DSPEs) such as Apache Flink, Storm, and Spark Streaming.
In addition, it will be extensible to work with future DSPEs, exposing flexible APIs for the definition of new data input and output types.

\subsection{S2CE Performant Interconnections}
\label{sec:apis}

The various modules of \pl\ will be interconnected through two standardized APIs, as shown in Figure \ref{fig:apis}.
These interfaces will be high-performant, secured, capable of exchanging voluminous and fast, exascale data, as well as rich data, depending on the modules consuming them.

The \textit{External API} will be responsible for consuming data inputted to the platform from various upstream sources (other platforms or end-user devices), or outputted to downstream platforms and devices for consumption (further analysis, visualization, and storage). 
The \textit{Internal API} will be responsible for allowing the various modules to interact and exchange raw data, tuned parameters, models, etc.
For example, the Cloud and Edge Management components will tune the computing infrastructure for a given data stream and model applied, and provide such tunings to the Input Interface for adjusting sampling rates.
Furthermore, the Transformations and Machine Learning components will consume raw streams or pre-models already built at the edge, to build final, full-blown ML models.
Performance metrics will be used to monitor and adjust resource consumption both at the cloud and edge infrastructure.

These APIs allow \pl\ to have the following key features with respect to big data mining systems:
\begin{itemize}[leftmargin=*]
	\item High extensibility and flexibility to accommodate future heterogeneous data sources and pools, distributed computing platforms, ML algorithms, tuning algorithms, etc.
	\item Scalability to consume data from multiple sources with extreme volumes and velocity.
	\item Federated-ness for connecting and building big data stream pipe\-lines across different clouds and edge.
	\item Standardized interoperability with other stream or batch platforms, for faster and efficient data sharing.
\end{itemize}

\section{Innovation Potential}
\label{sec:innovation-potential}

The \pl\ platform will enable the execution of machine and deep learning algorithms on a variety of existing (and future) cloud providers and DSPEs that are widely used in industry.
This section outlines the key innovation potential of the \pl\ platform in various domains.

\subsection{Cloud Provisioning and Orchestration}

The cloud landscape is becoming quite complex with different cloud providers and offerings that cover the full spectrum from IaaS to PaaS and SaaS with several intermediate solutions~\cite{linthicum19cloud-complexity, gabarin19cloud-complexity2}.
Vendor choosing and careful capacity planning is nowadays inevitable, given the different prices, APIs, tools, and services that may vary from provider to provider~\cite{samuels18choose-cloud}.
Nowadays, this process is necessary given that once a provider is chosen, many workflows such as provisioning, fault tolerance, development, DevOps, QA, etc., are consequently tied to the offerings of the chosen provider.
In fact, all the aforementioned dependencies make the process of changing providers or even collaboration of processes between different providers a very difficult task~\cite{martins16vendor-lockin}.
It is then necessary to develop solutions that may encompass the provisioning of resources and orchestration of software across different providers in a vendor-neutral manner.
That way, splitting processes across several providers or even changing from one to another, becomes a much easier task.

When it comes to data processing and ML, it is even more important, given the myriad of software that is available and may or may not collide with the different managed solutions available in the major cloud providers such as Amazon, Microsoft, and Google~\cite{samuels18choose-cloud}.
Many enterprises might go for a single product based on a single platform just because of the convenience of such managed solutions, getting into a game of vendor lock-in that makes it difficult to migrate to any other, or even to run some tasks on premises~\cite{martins16vendor-lockin}.
\pl\ can be easy to install and manage, as well as optimize resource allocation, software deployment, and runtime tuning across different cloud providers in a vendor-agnostic way.

\subsection{Edge Preprocessing and Movement}

Given the heterogeneous nature of the Edge, resource optimization and task orchestration is an issue that is difficult to tackle in a single way~\cite{zhang19heteroedge, schafer16tasklets, varghese16challenges-edge}.
Advances in execution paradigms such as containers and container-orchestration software may make it easier~\cite{hoque17containers-orchestration}, but given the difference in architecture and computation or storage capacity that any node in an Edge infrastructure might have, task placement strategies and resource monitoring become essential tasks to guarantee the correct execution of preprocessing software~\cite{shi16edge-vision, skarlat17qos-fog, stream-edge-survey18}.

Even with that, especially in streaming data analytics, the volume of ingestion might grow unexpectedly, overloading nodes that were not meant to deal with such a volume of incoming data.
In some cases, scalability can be achieved when underloaded nodes can collaborate in distributed tasks, or even between different edge infrastructures closely collocated.
However, there is a point in which a mechanism is needed to offload this workload to the cloud, where computation power might grow as needed.
\pl\ will provide task placement mechanisms and algorithms across edge nodes, as well as offloading methods for tasks to be migrated to the cloud from edge, seamlessly and without impacting performance or latency beyond agreed SLAs.

\subsection{Data-driven Decisions based on Streams}

Big Data projects, to be truly beneficial for organizations, have to provide not only data storage and retrieval capabilities, but also support for data-driven business decisions, inevitably necessitating data analytics tools~\cite{19cloud-analytics, 17cloud-analytics-2}.
In the case of big data resources, due to little or no a-priori knowledge on inter-dependencies present in the data, ML techniques are of particular use.
In a traditional setting, they are executed on a regular basis in batch mode.
However, in many industry cases nowadays, a tool is needed for advanced analytics on data streams coupled with ML, and performed in near-real time.
Thus, it is of no surprise that this need for online analytics raised the interest in stream processing engines from major industrial players~\cite{hecht19cloud-stream} (e.g., IBM, Twitter, LinkedIn, Google, Amazon).
\pl\ will provide exactly such analytics to support exascale data-driven decisions on business and innovation.

\subsection{Innovation Exchange with Apache Ecosystem's Open Source}

The big data interest was followed by a rapid development of DSPEs, such as Apache Storm, Samza, and more recently Apache Spark and Apache Flink, which are the dominating solutions in the area of online analytics.
These open source products achieved major popularity, exceeding the popularity of commercial, closed solutions.
This is clearly confirmed by the offerings of data processing vendors such as Oracle providing its Oracle Table Access for Hadoop and Spark components~\cite{oracle19hadoop-spark}.
Other vendors, such as SAS Institute and IBM, developed their own stream-related offerings, such as the SAS Cloud Analytics~\cite{sas19cloud-analytics} and IBM Streams~\cite{ibm19stream-analytics}, respectively.
Some of these systems can or even are advised to be blended with open source ML solutions including Spark MLLib~\cite{ibm19stream-analytics-2}.
In addition, commercial offerings from Cloudera and other companies integrating big data projects into comprehensive platforms have provided support and integration with these dominant streaming engines and analytics.

The risks introduced by developing new solutions, and major costs of big data projects arising from hardware infrastructure are commonly observed.
Hence, many companies build their big data solutions based on open source platforms yielding no license cost, or costs for core development of the tools.
This is clearly confirmed by the industrial success of Apache big data projects.
Naturally, \pl\ must integrate organically with the Apache ecosystem projects as a next generation big data solution, and enable innovation exchange with existing and future industrial solutions.

\subsection{Unifying Efforts for a Stream ML Library}

\begin{figure}[t]
\begin{center}
	\includegraphics[scale=0.4]{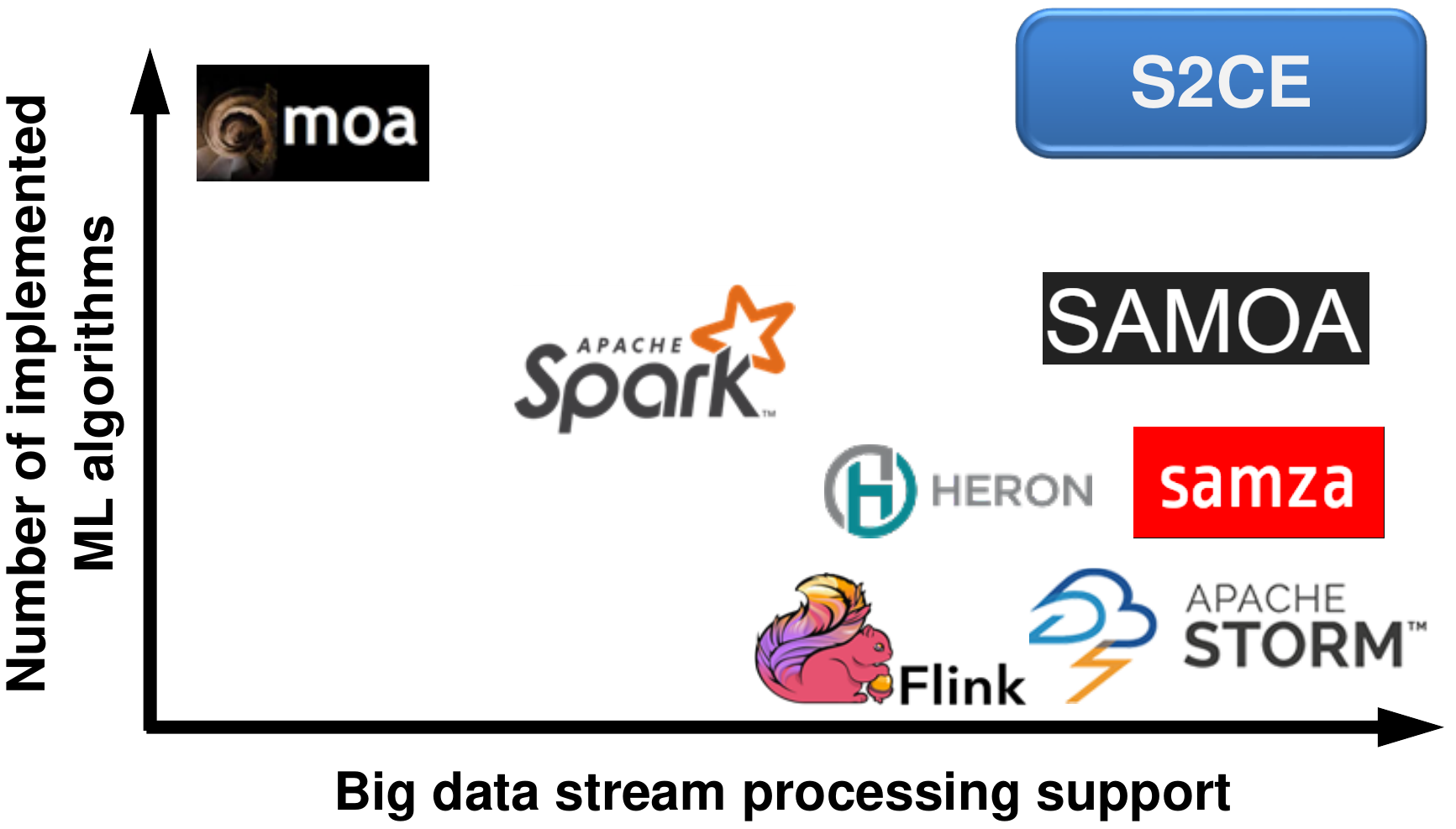}
	\caption{\pl\ positioning in the big data stream processing ecosystem.}
	\Description[]{\pl\ positioning in the big data stream processing ecosystem.}
	\label{fig:comparison}
\end{center}
\vspace{-4pt}
\end{figure}

Many real life big data use cases demand the inclusion of ML techniques in data processing.
This requirement was partly answered by open-source libraries extending the functionality of Apache Storm, Spark, and Flink~\cite{karimov18sdspe-comparison}.
Unfortunately, major deficiencies of this approach can be already observed.
The effort of the open source community is fragmented across different DSPE projects.
As the popularity of DSPEs changes significantly over relatively short time periods depending on industrial support, there is a risk that none of the ML libraries accompanying nowadays DSPEs will reach its maturity, before DSPEs are replaced by a competing and more complete framework~\cite{prakash18dspes-adoption}.
To the point, the industry focus already gradually moved from Apache Storm to Spark, and lately, DSPEs such as Flink are appreciated and significant developing effort is applied.
Importantly, none of the ML libraries accompanying these DSPEs provide key stream mining techniques developed in the research community such as classification, clustering, or regression.

Hence, on the one hand, there is a growing gap between the constantly expanding portfolio of stream mining techniques in research projects such as MOA~\cite{moa-jmlr10}, and the limited availability of them in DSPEs adopted by the big data industry.
On the other hand, projects such as MOA include extensive implementations of state-of-the-art stream ML techniques.
However, being a research project, MOA does not include distributed processing support and integration interfaces or supervision interfaces, which are mandatory for industrial use.
\pl\ envisions to unify efforts in the open source space to produce a comprehensive ML library for big data stream mining (see Figure~\ref{fig:comparison} for a comparison).

\section{Concluding Remarks}
\label{sec:conclusion}

This paper makes the case for the need of an exascale data stream mining platform, that can address the current and future industrial challenges in big data processing.
Such a platform must scale seamlessly between available cloud and edge infrastructures to tackle the ever increasing volume, velocity, variety, and veracity of data expected in the future, and can compute complex machine and deep learning models needed for future, data-driven business decisions.
The paper proposes an architectural design for \pl, a platform that addresses all these needs and industrial challenges in a one-stop-shop solution.

\pl\ is a first of its kind, optimized, multi-cloud and edge orchestrator, easily configurable, scalable, and extensible, while utilizing cloud and edge smart resource management and distributed processing.
\pl\ does not need to be linked to a single cloud provider or DSPE.
It has been inspired by the various big data projects in the Apache ecosystem.

The author team has already started the design and development of the needed components, by bootstrapping on functionalities of Apache SAMOA.
\pl\ will be developed and continuously refined, irrespective of evolving popularities of existing and future DSPEs and cloud infrastructures.

Moreover, \pl\ aims to bridge the gap between industry-selected solutions and research projects by offering innovative approaches to stream mining through:

\begin{enumerate}[leftmargin=*, itemsep=1pt]
	\item The ability to optimize the use of available cloud and edge resources by deploying stream mining tasks using existing DSPE clusters preferred by individual organizations.
	\pl\ can go beyond current approaches by automatically tuning the platform, the available DSPEs and cloud (public, private, or hybrid) infrastructure, breaking down the streaming processing task at hand into subcomponents for efficient execution and minimization of computing resources by offloading essential preprocessing to the edge.
	Importantly, this will let the research community to contribute to the unified ML library of \pl, rather than multiple native libraries of different DSPEs.
	Moreover, the platform will include algorithms estimating the performance and DSPE settings suitable for ML.

	\item Extensive set of high-impact tools on the ultimate results of ML process, including: (i) stream sampling methods to control the data volume used to update ML models in order to prevent performance issues; (ii) synthetic stream generation methods to preserve privacy and confidentiality of true data, while unlocking the value of otherwise frequently not used or available data streams; (iii) a visualization module revealing inference rules present in ML models, making them transparent and promoting business use of stream mining via increased understanding of the performance of models.

	\item The adoption of existing industry standards.
	In particular, the platform can provide integration components with popular systems such as Kafka, making it possible to create efficient and high throughput data pipelines using \pl.
	The \pl\ should be provided together with thoroughly designed integration patterns for using \pl\ with third-party modules such as data storage platforms (e.g., Apache HBase), data ingestion (e.g., Apache Flume), and DSPEs (e.g., Apache Storm, Apache Flink).
	Finally, \pl\ can be coupled with extensive monitoring and management abilities, easing \pl\ adoption across companies.
\end{enumerate}

\smallskip
\noindent
In conclusion, we envision \pl\ to become the go-to, one-stop-shop platform for mining of big data streams over cloud and edge, and adopted by big data practitioners for its easy pipeline integration and management, and extended by machine learning experts.

\balance


\bibliographystyle{ACM-Reference-Format}
\bibliography{main}  

\end{document}